\documentstyle[12pt,worldsci,psfig,epsfig]{article}
\begin{document}
\title{HEAVY QUARK CONFINEMENT FROM DYSON-SCHWINGER EQUATIONS}
\author{Conrad Burden\\
{\em Department of Theoretical Physics,}\\
{\em Research School of Physical 
  Sciences and Engineering,}\\
{\em Australian National University, Canberra ACT 0200, AUSTRALIA}}
\vspace{0.3cm}
\maketitle
\setlength{\baselineskip}{2.6ex}

\vspace{0.7cm}
\begin{abstract}
A formalism for studying the confinement of heavy quarks by
considering the renormalised quark Dyson-Schwinger equation in the limit
$m_Q \rightarrow \infty$ is described.  We are particularly interested 
in studying the analytic structure of heavy quark 
propagators in the region of the
complex momentum plane close to the bare fermion mass pole.

\end{abstract}
\vspace{0.7cm}

\section{Introduction}

Confinement and chiral symmetry breaking in quantum chromodynamics (QCD)
must ultimately have a field theoretic cause.  They are innately 
nonperturbative phenomena, demanding the use of appropriate 
techniques if they are to be fully understood.  Lattice gauge theory and the 
Dyson-Schwinger (DS) equation technique\cite{RW94} represent the two most 
promising lines of attack on QCD from a nonperturbative, field theoretic 
point of view. 

Of these two, lattice gauge theory has the advantage that it is as close 
as one can find to an {\it ab initio} treatment of QCD.  Approximations 
are clearly defined and in principle controllable.  It has the disadvantage 
that a qualitative understanding of physical phenomena tends to be obscured 
by numerical brute force.  Some progress is now being made, 
however, in interpreting lattice gauge fields in terms of monopoles and 
flux tubes \cite{GB98}.  

Dyson-Schwinger equations provide a complementary approach to QCD.  The 
infinite tower of coupled integral equations relating Greens functions 
is an exact, nonperturbative description of QCD.  Before any progress can 
be made, the tower must be truncated, generally at the level of quark and 
gluon propagator equations and bound state Bethe-Salpeter and Faddeev 
equations.  Chiral symmetry breaking is signalled by the generation of 
a non-zero scalar part to the quark propagator, while a sufficient condition 
for quark confinement is the absence of real, timelike singularities in the 
propagator.  These conditions ensue provided the dressed gluon propagator 
is sufficiently infrared enhanced \cite{R98}.  

Numerical solution of the gluon propagator DS equation is a 
formidable task, and opinions differ about the resultant infrared behaviour 
of the gluon propagator \cite{P96,AB98}.  It has been argued however that, 
if the three gluon and ghost-gluon vertices are modelled in a way free of 
singularities, a strongly infrared enhanced gluon propagator may result, 
leading to the required chiral symmetry breaking and quark 
confinement \cite{HMR98}.  Within the DS picture, these 
two phenomena go hand in hand, both driven by the functional form of 
the quark propagator.  

There is of course no necessary inconsistency between 
the DS picture and dual superconductor flux tube models.  
A strongly infrared enhanced gluon propagator can readily generate a 
Wilson area law \cite{W82}.  A fortiori,  one should not be perturbed that 
the flux tube picture is not immediately manifest in the interpretation 
of a timelike singularity free propagator as a signal of confinement.  
After all, that Wilson loops should satisfy an area law (which we interpret 
as the existence of a flux tube) is not an immediate, or even necessary 
consequence of the fact that Polyakov loops are zero valued (which we 
interpret as a signal of colour confinement).  Obviously, 
both pictures of confinement can happily coexist.  

In what follows we give a very brief description of a formalism designed 
to extend the DS picture to the realm of heavy quarks 
\cite{BL97,B98,B98a}.  We shall see that the DS picture of 
quark confinement can be carried over to the extreme heavy quark limit 
in a natural and instructive way.  

\section{The Heavy Quark Limit}

Our starting point is the quark DS equation 
\begin{eqnarray}
S(p)^{-1} & = & i \gamma\cdot p A(p^2)  + B(p^2) \\
          & = & Z_2[ \gamma\cdot p + m_0(\Lambda)] + 
   Z_1\int^\Lambda \frac{d^4q}{(2\pi)^4} g^2 D_{\mu\nu}\gamma_\mu
              S(q)\Gamma_\nu(q,p), 
\end{eqnarray}
with $S$, $D_{\mu\nu}$ and $\Gamma_\mu$ the renormalised dressed quark 
propagator, gluon propagator and proper quark-gluon vertex respectively.  
The renormalisation scale $\mu$ is set such that 
$\left.S(p)\right|_{p^2 = \mu^2} = 1/[i\gamma\cdot p + m_R(\mu^2)]$.
We use a Euclidean metric in which timelike vectors satisfy 
$p^2 = -p_{\mathrm Minkowski}^2 < 0$, 
and for which $\{\gamma_\mu,\gamma_\nu\} = 2\delta_{\mu \nu}$.  

The heavy quark limit is taken by assuming the expansions 
\begin{equation}
A(p^2,\mu^2) = 1 + \frac{\Sigma_A(K,\kappa)}{m_R(\mu^2)} + \ldots,  
\hspace{5 mm}
B(p^2,\mu^2) = m_R(\mu^2) + \Sigma_B(K,\kappa) + \ldots,  \label{sigBdef}
\end{equation}
where we have defined the momentum variable $K$ 
and renormalisation point $\kappa$ by 
\begin{equation}
K = \frac{p^2 + m_R^2}{2im_R}, 
\hspace{5 mm}
\kappa = \frac{\mu^2 + m_R^2}{2im_R}. \label{Kdef}
\end{equation}
The change of independent variable 
$p^2 \rightarrow K$ induced by the transformation Eq.~(\ref{Kdef}) 
is discussed in detail in reference~ \cite{B98} .  The main point to 
note here is that the real timelike $p^2$ axis in the vicinity of the 
`would-be bare fermion mass pole', $p^2 = -m^2$, becomes the imaginary 
$K$ axis in the new coordinates.  The aim of the exercise is to solve 
for the quark propagator in the vicinity of the bare mass pole $K = 0$.  

\begin{figure}[t]
\begin{centering}
\psfig{figure=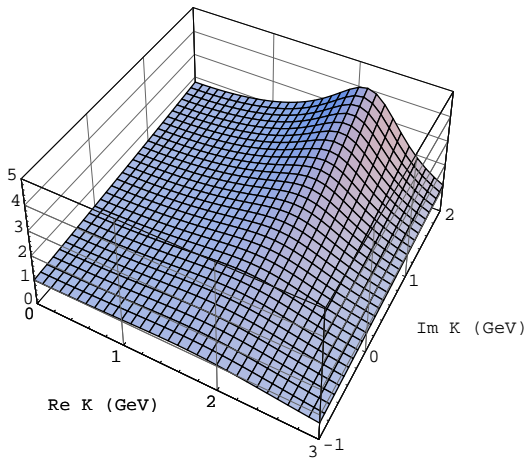,height=100 mm}
  \caption[x]{The modulus $\left|\sigma_Q(K)\right|$ of the heavy quark 
propagator in the complex $K$ plane obtained by solving the heavy quark 
DS equation in bare vertex approximation with the gluon propagator 
$\Delta(k^2)$.}  
\label{fig1}
\end{centering}
\end{figure}

For the purposes of expanding the DS equation in the inverse heavy quark 
mass we introduce the momentum substitutions generic to heavy quark effective 
theory, 
\begin{equation}
p_\mu = im_R v_\mu + k_\mu, 
\end{equation}
where $v_\mu v_\mu = 1$.  This gives $K = k\cdot v + k^2/(2im_R)$, 
and so in the vicinity of $k_\mu = 0$, or $K=0$, the dressed quark 
propagator has the form, 
\begin{eqnarray}
S(p,\mu) & = & \frac{1 + \gamma\cdot v}{2} 
       \frac{1}{iK + \Sigma_B(K,\kappa) - \Sigma_A(K,\kappa)} 
    + O\left(\frac{1}{m_R}\right) \nonumber \\
      & \stackrel{\mathrm{def}}{=} 
        & \frac{1 + \gamma\cdot v}{2} \sigma_Q(K,\kappa) 
    + O\left(\frac{1}{m_R}\right).     \label{hprop}
\end{eqnarray}
The confinement criterion that $S(p)$ should be free from timelike 
singularities on the negative real $p^2$ axis translates in the heavy 
quark case to a requirement that $\sigma_Q$ should be free from 
singularities on the imaginary $K$ axis.  

In reference~ \cite{B98a} the heavy quark propagator is solved for using 
the bare vertex or ``rainbow'' approximation $\Gamma_\mu(q,p) = \gamma_\mu$, 
and Landau gauge smeared Frank and Roberts\cite{FR96} model gluon 
propagator $D_{\mu\nu}(k) = (\delta_{\mu\nu} - k_\mu k_\nu/k^2) \Delta(k^2)$, 
where 
\begin{equation}
\Delta(k^2) = (2\pi)^4 \frac{m_t^2 d}{\alpha^2 \pi^2}
      e^{-k^2/\alpha} + 4\pi^2 d \frac{1 - e^{-k^2/(4m_t^2)}}{k^2}.  
                                          \label{GFRprop}
\end{equation}
Here $d = 12/(33 - 2N_f)$, $N_f = 3$ is the number of light quark 
flavours, $m_t = 0.69$ GeV is a parameter fitted to a range 
of calculated pion observables, and the Gaussian width $\alpha$ is chosen 
to be 0.5643 (GeV)$^2$.  

In Fig.~\ref{fig1} is plotted the modulus 
$\left|\sigma_Q(K)\right|$ of the heavy quark propagator 
obtained by solving the DS equation in the vicinity of $K = 0$, with 
a renormalised heavy quark mass of $m_R = 5$ GeV.  We see that the bare 
propagator pole at $K = 0$ has moved well away from the imaginary 
$K$ axis, leaving the heavy quark propagator analytic over a large region 
in the vicinity of $K = 0$.  The removal of propagator poles from the 
imaginary $K$ axis signals heavy quark 
confinement. As noted in reference~ \cite{B98} , a large shift of the pole 
portends well for the possibility of finding solutions 
to the Bethe-Salpeter equation for $Q$-$\bar{q}$ mesons\cite{BL97}.  
Solution of the heavy meson Bethe-Salpeter equation, of which accurate 
determination of heavy quark propagators is a necessary first step, should 
prove helpful in impulse approximation studies of semileptonic 
decays\cite{IKMR98}.

\section*{Acknowledgement}

The author acknowledges helpful discussions with F.\ T.\ Hawes and 
C.\ D.\ Roberts, and the hospitality of the Special Research Centre 
for the Subatomic Structure of Matter in Adelaide for hosting the 
Workshop on Nonperturbative Methods in Quantum Field Theory where part 
of this work was completed.

\end{document}